\documentclass[preprint,showpacs,preprintnumbers,amsmath,amssymb]{revtex4}

\usepackage{graphicx}
\usepackage{dcolumn}
\usepackage{bm}

\def\slash#1{\not\!#1}

\def\delsla{\!\!\not\!\partial}

\begin{document}

\preprint{OCU-PHYS 269}

\title{Phase diagram of quark-antiquark and diquark condensates \\
 at finite temperature and density \\
 in the 3-dimensional Gross Neveu model }

\author{Hiroaki Kohyama}
 \email{kohyama@sci.osaka-cu.ac.jp}
\affiliation{%
Department of Physics, Osaka City University, 
Sumiyoshi-ku, Osaka 558-8585, JAPAN
}%

\date{\today}

\begin{abstract}
We construct the phase diagrams of the quark-antiquark and diquark condensates
at finite temperature and density 
in the 3D (dimensional) 2-flavor Gross Neveu model.
We found that, in contrast to the case of the 4D Nambu Jona-Lasinio model,
there is no region where the quark-antiquark and diquark condensates coexist.
The phase diagrams obtained 
for some parameter region show similar structure with the 4D QCD phase diagram.
\end{abstract}

\maketitle

\section{\label{sec:level1}Introduction}
The four-fermion interaction models are considered to be the effective theory for describing
phase transitions. In $3+1$ dimensions (4D), the Nambu Jona-Lasinio (NJL) model 
can be regarded as an effective theory of QCD\cite{NJL}.
The model successfully reproduces the chiral phase transition in quark matter
(see, e.g. \cite{HatsuKuni,onNJL}).
Furthermore, through recent studies of the diquark condensate, a variety of color 
superconducting phases are expected to be realized\cite{Alford}.
It has been shown that there is the region where the chiral(quark-antiquark) and diquark
 condensates coexist. (For a nice review, see, e.g. \cite{Mei}.)

The Gross-Neveu (GN) model proposed in 1974, is a model of Dirac fermions 
interacting via four-fermion interactions\cite{GN}.
The 2D GN model is a renormalizable quantum field 
theory, and the 3D GN model is renormalizable in the leading $1/N$ order. They are
closely related to the Bardeen-Cooper-Schrieffer
theory of superconductivity\cite{BCS}.
Although the lower dimensional theory seems not
to be realistic, the 2D GN type models are believed 
to be the effective models of 1 dimensional
condensed matter systems such as conducting polymers like polyacetylene\cite{polymer}. 
Since the GN model shares many properties with 
QCD, notably asymptotic freedom, chiral symmetry breaking in vacuum,
the comparison of QCD and the GN model is an
interesting subject.

A variety of works has been devoted to the study of the GN 
model\cite{onGN,Mal,Kanemura,Ulli,Klimenko,Kneur,Repre,Zhou}.
By using bag-model boundary conditions, a closed
formula for the effective renormalized coupling constant in the large-N limit, 
was derived in \cite{Mal}.
In a constant curvature space, 
the phase structure of chiral symmetry breaking in the GN
model at finite temperature and density is discussed in \cite{Kanemura}.
The phase diagrams of the quark-antiquark condensate in the 2D GN model 
was obtained in \cite{Ulli}.
In 3D, analyses so far have been done
choosing 2-dimensional(2d) or 4-dimensional(4d) spinor representations for quarks.
The case of the 2d representation is interesting itself because it is the nontrivial
lowest-order representation. The phase diagram of the quark-antiquark condensate
was constructed by employing the 
2d representation in \cite{Klimenko}.
On the other hand in the case of the 4d representation,
there exists the $\gamma^5$ (see, e.g. \cite{Appel}) and the properties of the model
bear resemblance to the NJL model in 4D. In this sense, this case is also interesting
and the phase diagram of the quark-antiquark condensate was obtained
in \cite{Kneur}.
The relation between the 2d and 4d representations are discussed in 
\cite{Repre}.
Furthermore, through analyzing the 3D GN with the 2d representation, 
the phase structure of the quark-antiquark($\bar{q}q$) and diquark($qq$)
condensates in vacuum (zero temperature and chemical potential($T=\mu=0$))
was studied in \cite{Zhou} under the condition that the $\bar{q}q$ and $qq$ 
condensates are much smaller than the cut-off scale of the model. It was found that
the $\bar{q}q$ and $qq$ condensates do not coexist at $T=\mu=0$.

In this paper, we study the $\bar{q}q$ and $qq$ condensates at finite 
temperature and density in the 3D GN model with the 2d representation. We obtain the 
phase diagrams, and discuss the similarities and differences 
between the 3D GN model and the 4D NJL model.

The plan of the paper is as follows: In Sec.II we present the Lagrangian of the 3D GN model 
and introduce the mean-field approximation.
In Sec.III we derive the thermodynamic potential. In Sec.IV
we present the results of the numerical analyses.
Among others, we show that there is no region where the $\bar{q}q$ and $qq$ condensates coexist. 
In Sec.V we display the phase diagrams. We find 
that the structure of phase diagrams bear resemblance to the QCD phase diagram for 
some parameter region. Sec.VI
is devoted to summary and conclusions. In Appendixes A and B we describe the intermediate calculation
and the renormalization
for the thermodynamic potential.

\section{Gross Neveu Model}
The general form of the Lagrangian density of the 3D 2 flavor massless Gross Neveu model 
with 2d representation reads
\begin{align}
\mathcal{L} = \bar{q}i \delsla q & + G_S [(\bar{q}q)^2 + (\bar{q}\vec{\tau}q)^2] \nonumber \\
    & {}+ G_D \sum_{a=1}^8(\bar{q} \vec{\tau} \lambda_a q^C) (\bar{q}^C \vec{\tau} \lambda_a q)
\label{geneGN}.
\end{align}
Here $\vec{\tau}=(\tau_1,\tau_2,\tau_3)$ are the Pauli matrices in flavor space and 
$\lambda_a$ is a Gell-Mann 
matrix in color space. 
$G_S$ and $G_D$ are the coupling constants of the $\bar{q}q$ 
and $qq$ interactions, respectively, and $C$ is the charge conjugation matrix. 
For the Dirac $\gamma$ matrices and 
$C$,  we use the forms as in \cite{Zhou},
\begin{eqnarray}
\gamma^0 = 
\left(
\begin{array}{cc}
1 & 0\\
0 & -1
\end{array}\right)\;,
\gamma^1 = 
\left(
\begin{array}{cc}
0 & i\\
i & 0
\end{array}\right)\;,
\gamma^2 = 
\left(
\begin{array}{cc}
0 & 1\\
-1 & 0
\end{array}\right) = C.
\end{eqnarray}
The charge conjugated field are defined by
\begin{equation}
q^C = C\bar{q}^T, \, \, \bar{q}^C=q^T C.
\end{equation}

Following the reasonings described in \cite{Mei}, we reduce the Lagrangian density in
Eq.(\ref{geneGN}) to
\begin{align}
\mathcal{L} = \bar{q}i \delsla q & + G_S (\bar{q}q)^2 \nonumber \\
    & {}+ \sum_{a=2,5,7} G_D (\bar{q} \tau_2 \lambda_a q^C) (\bar{q}^C \tau_2 \lambda_a q)
\label{preGN}.
\end{align}
$\mathcal{L}$ in Eq.(\ref{preGN}) enjoys various symmetry properties, which are fully
discussed in \cite{Zhou}.
Here we choose a color direction for diquark condensate to blue, which is equivalent
to select $\lambda_2$ in Eq.(\ref{preGN}) (see \cite{Mei}).
Then we finally arrive at the following Lagrangian density:
\begin{align}
\mathcal{L} = \bar{q}i \delsla q & + G_S (\bar{q}q)^2 \nonumber \\
    & {}+  G_D (\bar{q} \tau_2 \lambda_2 q^C) (\bar{q}^C \tau_2 \lambda_2 q)
\label{GN}.
\end{align}
Due to $\lambda_2$ in Eq.(\ref{GN}),
 only two colors (red, green) participate in the 
$qq$ condensate while all three colors (red, green, blue) do in 
the $\bar{q}q$ condensate.

Let us introduce the mean-field approximation and rewrite the Lagrangian density as follows:
\begin{align}
\mathcal{L} = & \bar{q}i \delsla q 
     -\bar{q} \sigma q 
    - \frac{1}{2}\Delta^{*}(\bar{q}^C \tau_2 \lambda_2 q) \nonumber \\
    &{} - \frac{1}{2}\Delta(\bar{q} \tau_2 \lambda_2 q^C) 
     -\frac{\sigma^2}{4 G_S} - \frac{|\Delta|^2}{4 G_D} \, ,
\label{mGN}
\end{align}
where $\sigma$ and $\Delta$ are the order parameters for
the $\bar{q}q$ and $qq$ condensates:
\begin{eqnarray}
    \sigma = -2G_S \langle \bar{q}q \rangle 
    \quad {\rm and} \quad
    \Delta = -2G_D \langle \bar{q}^C \tau_2 \lambda_2 q \rangle.
\end{eqnarray}

To deal with finite density system, we introduce a chemical potential
being conjugate to the quark number $\bar{q}\gamma^0 q$. Then the Lagrangian density reads
\begin{align}
\mathcal{L} = & \bar{q}i \delsla q + \bar{q} \mu \gamma^0 q
     -\bar{q} \sigma q 
    - \frac{1}{2}\Delta^{*}(\bar{q}^C \tau_2 \lambda_2 q) \nonumber \\
    &{}- \frac{1}{2}\Delta(\bar{q} \tau_2 \lambda_2 q^C)
     -\frac{\sigma^2}{4 G_S} - \frac{|\Delta|^2}{4 G_D}
\label{muGN}.
\end{align}
Introducing the Nambu-Gorkov basis \cite{Nambu} 
\begin{eqnarray*}
     \Psi = 
     \left(
     \begin{array}{c}
     q \\
     q^C
     \end{array}\right)
     \quad {\rm and} \quad
     \bar{\Psi} = 
     \left(
     \, \bar{q} \,\,\, \bar{q}^C
     \right) \, ,
\end{eqnarray*}
and using the relation $\bar{q}^C \gamma^0 q^C = -\bar{q} \gamma^0 q$,
we can write the Lagrangian density in a momentum space as
\begin{align}
\mathcal{L} &= 
    \frac{1}{2} \bar{\Psi} G^{-1} \Psi
     -\frac{\sigma^2}{4 G_S} - \frac{|\Delta|^2}{4 G_D},
\end{align}
where
\begin{eqnarray}
    G^{-1}  &=  \left(
    \begin{array}{cc}
       (\slash{p} - \sigma + \mu \gamma^0){\bf 1}_f {\bf 1}_c
     & -\tau_2 \lambda_2 \Delta{\bf 1}_s \\
       -\tau_2 \lambda_2 \Delta^{*}{\bf 1}_s
       & (\slash{p} - \sigma - \mu \gamma^0){\bf 1}_f {\bf 1}_c
    \end{array}\right)
\label{NGGN}.
\end{eqnarray}
${\bf 1}_f ,{\bf 1}_c$ and ${\bf 1}_s$ are the unit matrix in flavor, color
and spinor space respectively.

\section{The thermodynamic potential}
\subsection{Derivation of the thermodynamic potential}
Following the standard method, we can evaluate the thermodynamic potential:
\begin{align}
     \Omega & (\sigma,|\Delta|) = 
       \frac{\sigma^2}{4G_S} + \frac{|\Delta|^2}{4G_D} \nonumber \\
      &{} - \frac{1}{\beta V} \ln \int [d\Psi] \exp 
     \left[ 
     \frac{1}{2}\sum_{n,{\bm p}} \bar{\Psi} (\beta G^{-1}) \Psi 
     \right],
\label{thermo}
\end{align}
where $\beta = 1/T$ is the inverse temperature and $V$ is the volume of
the system.
With the help of the formula
\begin{align}
     \int [d\Psi] \exp 
     \left[ 
     \frac{1}{2}\sum_{n,{\bm p}} \bar{\Psi} (\beta G^{-1}) \Psi 
     \right]
     =
     {\rm Det}^{1/2} (\beta G^{-1}) \, ,
\end{align}
we can rewrite Eq.(\ref{thermo}) as
\begin{align}
     \Omega & (\sigma,|\Delta|) = 
      \frac{\sigma^2}{4G_S} + \frac{|\Delta|^2}{4G_D}
       - \frac{1}{\beta V} \ln {\rm Det}^{1/2} (\beta G^{-1}) \, .
\label{thermodet}
\end{align}
After some manipulations which is given in Appendix A, the determinant becomes
\begin{eqnarray}
    {\rm Det}^{1/2} (G^{-1})  
    &= \bigl[ p_0^2 - E_{\Delta}^{+\,2} \bigr]^2 
       \bigl[ p_0^2 - E_{\Delta}^{-\,2} \bigr]^2
       \bigl[ p_0^2 - E^{+\,2} \bigr] 
       \bigl[ p_0^2 - E^{-\,2} \bigr],
    \label{det}
\end{eqnarray}
where $p_0 = i(2n+1)\pi T ,\, (n=\cdots,-2,-1,0,1,2,\cdots)$ and
\begin{align}
 & E^\pm \equiv E \pm \mu \, , \quad 
   E \equiv \sqrt{\vec{p}^{\,\, 2} + \sigma^2} \, , \quad
   \vec{p}^{\,\, 2} = p_1^2 + p_2^2
\nonumber \\
&E_{\Delta}^\pm{}^2 \equiv E^2 + \mu^2 + |\Delta|^2 \pm 2 \sqrt{E^2 \mu^2 + \sigma^2 
|\Delta|^2} \,\,\, (\geq 0).
\end{align}
Thus, we obtain
\begin{align}
     \Omega & (\sigma,|\Delta|) = 
      \frac{\sigma^2}{4G_S} + \frac{|\Delta|^2}{4G_D} \nonumber \\
     & {} - T \sum_{\pm} \sum_n \int \!\! \frac{d^2 p}{(2\pi)^2}
     \biggl[
     \ln[\beta^2 (p_0^2 - E^{\pm}{}^2)] 
     + 2 \ln[\beta^2 (p_0^2 - E_{\Delta}^{\pm}{}^2)]      
     \biggr].
\label{Thermody}
\end{align}

The frequency summation may be performed in a standard manner\cite{LeBellac}:
\begin{equation}
 \sum_n \ln [\beta^2 (p_0^2 - E^2)]= \beta [ E + 2 T \ln( 1+e^{-\beta E} ) ].
\end{equation}
Then we finally obtain
\begin{align}
     \Omega  (\sigma,|\Delta|) &= \Omega_{0}(\sigma,|\Delta|)+ \Omega_{T}(\sigma,|\Delta|),\\
     \Omega_0  (\sigma,|\Delta|) &= 
      \frac{\sigma^2}{4G_S} + \frac{|\Delta|^2}{4G_D}
      -2  \int \!\! \frac{d^2 p}{(2\pi)^2}
     \bigl[
     E + E_{\Delta}^+ + E_{\Delta}^-
     \bigr], \label{omega0} \\
     \label{Thermo2d}
     \Omega_T  (\sigma,|\Delta|) &= 
      -2 T \sum_{\pm} \int \!\! \frac{d^2 p}{(2\pi)^2}
     \biggl[
        \ln (1+e^{-\beta E^{\pm}}) + 2 \ln (1+e^{-\beta E_{\Delta}^{\pm}})
     \biggr].
\end{align}
Here $\Omega_0$ is $T$ independent contribution, which is ultraviolet divergent, while the
temperature dependent part $\Omega_T$ is finite.
For the purpose of later use, we write $\Omega_0$, in Eq.(\ref{omega0}), in the
integral form in the 3D Euclidean momentum space,
\begin{align}
     \Omega_0 & (\sigma,|\Delta|) = 
      \frac{\sigma^2}{4G_S} + \frac{|\Delta|^2}{4G_D} \nonumber \\
     & {} - \sum_{\pm} \int \!\! \frac{d^3 p_E}{(2\pi)^3}
     \biggl[
     \ln \Bigl( \frac{p_{E0}^2 + E^{\pm}{}^2}{p_E^2} \Bigr) 
     + 2 \ln \Bigl( \frac{p_{E0}^2 + E_{\Delta}^{\pm}{}^2}{p_E^2} \Bigr)
     \biggr].
\label{thermo03d}
\end{align}
The $p_E^2$ terms in the denominators in the second line of Eq.(\ref{thermo03d})
are inserted so as to drop an irrelevant infinite constant.

\subsection{Renormalized thermodynamic potential}
As mentioned in the previous subsection, $\Omega_0$, $T=0$ part of $\Omega$ 
is ultraviolet
divergent. To eliminate the divergences, we introduce the counter Lagrangian 
as derived in \cite{Klimenko} (see also \cite{GN}):
\begin{align}
 \mathcal{L}_C & = -\frac{1}{2} Z_S \sigma^2 - Z_D |\Delta|^2, \label{counterL}\\
 Z_S &= \frac{6}{\pi^2}\Lambda - \frac{3}{4} \alpha , \quad
 Z_D = \frac{2}{\pi^2}\Lambda - \frac{1}{4} \alpha,
 \label{counter}
\end{align}
where $\Lambda$ is the $3$D momentum cut-off and $\alpha$ is the arbitrary
renormalization scale. For completeness, the derivation of 
Eq.(\ref{counter}) is given in Appendix B.

Introducing the above counter Lagrangian, $\Omega_0$ in Eq.(\ref{thermo03d})
turns out to be finite and the renormalized $\Omega_{0r}$ becomes
\begin{align}
     \Omega_{0r} & (\sigma,|\Delta|) = 
      \Bigl( \frac{1}{4G_S}- \frac{3}{8}\alpha \Bigr) \sigma^2
      + \Bigr( \frac{1}{4G_D} - \frac{1}{4}\alpha \Bigr) |\Delta|^2 \nonumber \\
     & {} - \sum_{\pm} \int \!\! \frac{d^3 p_E}{(2\pi)^3}
     \biggl[
     \ln \Bigl( \frac{p_{E0}^2 + E^{\pm}{}^2}{p_E^2} \Bigr) 
     + 2 \ln \Bigl( \frac{p_{E0}^2 + E_{\Delta}^{\pm}{}^2}{p_E^2} \Bigr) 
      -\frac{3}{p_E^2} \sigma^2 -\frac{2}{p_E^2} |\Delta|^2
     \biggr].
\label{rezero}
\end{align}
Note that the counterterms cancel the divergences and the integral becomes finite.
Thus, after performing the renormalization, the renormalized thermodynamic potential
$\Omega_r$($\equiv \Omega_{0r} + \Omega_T$)
is finite and we carry out the numerical analyses on $\Omega_r$.

Before studying the $\bar{q}q$ and $qq$ condensates, we rewrite $\Omega_{r}$ 
by using the following parameters:
\begin{align}
\sigma_0 &\equiv -\frac{2\pi}{3}\Bigl( \frac{1}{4G_S} - \frac{3}{8}\alpha \Bigr), \label{sigma0}\\
\Delta_0 &\equiv -\pi \Bigl( \frac{1}{4G_D} - \frac{1}{4}\alpha \Bigr). \label{delta0}
\end{align}
Using $\sigma_0$ and $\Delta_0$, we can write $\Omega_r$ at $T=0=\mu$ as
\begin{align}
     \Omega_r&(\sigma,|\Delta|)\bigr|_{T=0=\mu} =
      -\frac{3}{2\pi}\sigma_0 \sigma^2 
      -\frac{1}{\pi}\Delta_0 |\Delta|^2 \nonumber \\
      &{} + \frac{1}{3\pi}\sigma^3
      + \frac{1}{3\pi}(\sigma+|\Delta|)^3 + \frac{1}{3\pi}|\sigma-\Delta|^3.
\label{zerozero}
\end{align}
Throughout in the following, we use the parameters $\sigma_0$ and $\Delta_0$.
By minimizing Eq.(\ref{zerozero}) for $\Delta=0$, one can easily verify that,
when $\sigma_0>0$, 
$\sigma_0$ corresponds to the $\bar{q}q$ condensate in vacuum.

\section{Quark-antiquark and diquark condensates}
We have obtained the thermodynamic potential in the previous section.
Eqs.(\ref{rezero}), (\ref{sigma0}) and (\ref{delta0}) 
tell us that this model has two free parameters ($\sigma_0,\Delta_0$).
In this paper, we aim to study the system where the $\bar{q}q$ condensate always takes
place in vacuum if $\Delta=0$, so we assume $\sigma_0$ to be positive (see, e.g.\cite{Ulli}).
After fixing $\sigma_0$, there remains a free parameter $\Delta_0$ and we 
introduce $r$ through
\begin{equation}
    r \equiv \Delta_0/\sigma_0.
\end{equation}
There is no direct way of fixing the parameter $r$, and we analyze for different $r$'s.

\begin{figure*}
\begin{center}
\includegraphics[width=16cm,height=16cm]{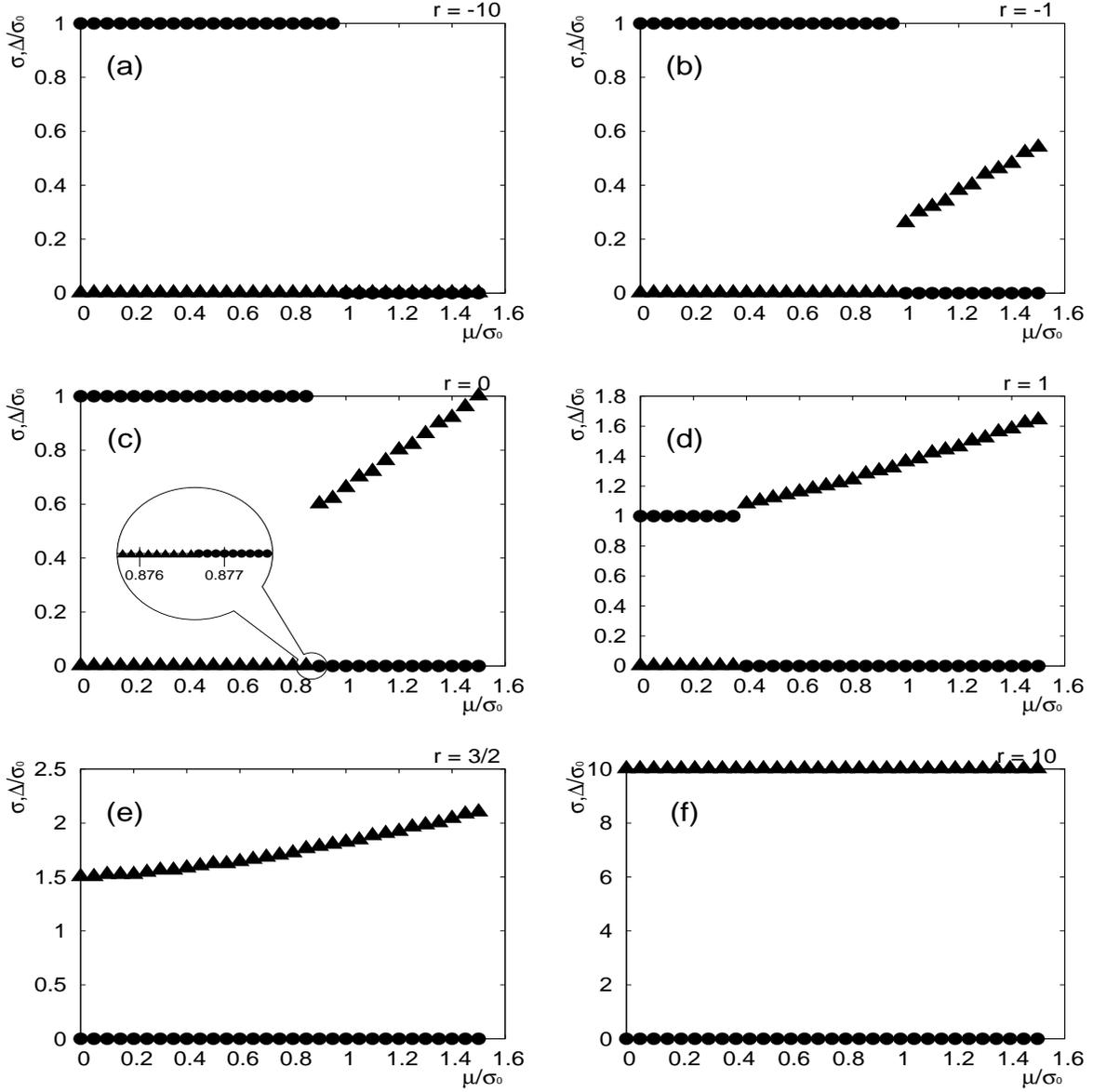}
\end{center}
\caption{\label{fig:0condensate}$\sigma$ (circles) and $\Delta$ 
  (triangles)
  as a function of chemical potential $\mu$ at $T=0$.}
\end{figure*}
Fig.~\ref{fig:0condensate} plots
the $\bar{q}q$ and $qq$ condensates (normalized by $\sigma_0$) at $T=0$. 
In the case of $r=-10$ (panel (a)),
we see that the $\bar{q}q$ condensate disappears at 
$\mu = 1.0 \sigma_0$, and there does not arise the $qq$ condensate.
Through numerical analyses, we have found that this is the case
for $r < -6.3$.
On the other hand for $r = -1$ (panel (b)), at $\mu = 1.0\sigma_0$ the $\bar{q}q$ condensate 
disappears and at the same time, the $qq$ condensate arises.
Similar results are obtained for $r=0$ and $1$, where the transition densities are
$\mu = 0.88\sigma_0$ and $\mu = 0.4\sigma_0$, respectively.
All the phase transitions in the panels (a)-(d) are of the first order.
With increasing the ratio $r$, 
the $qq$ condensate $\Delta$ becomes larger and eventually exceeds the
$\bar{q}q$ condensate at $\mu = 0$. As seen from the panel (e) for
$r = 3/2$, the $\bar{q}q$ condensate disappears and only the $qq$ condensate exists
for whole $\mu$.
More detailed analyses show that the
$\bar{q}q$ condensate does not occur for $r > 1.15$.
Thus the results are sensitive to 
the ratio $r=\Delta_0/\sigma_0$. 
It should be emphasized that
there is no region where the $\bar{q}q$ and $qq$ condensates coexist.
To clarify this fact, we show the close-up 
of the condensates near the phase transition point
in Fig.~\ref{fig:0condensate}(c).

\begin{figure*}
\begin{center}
\includegraphics[width=16cm,height=5.3cm]{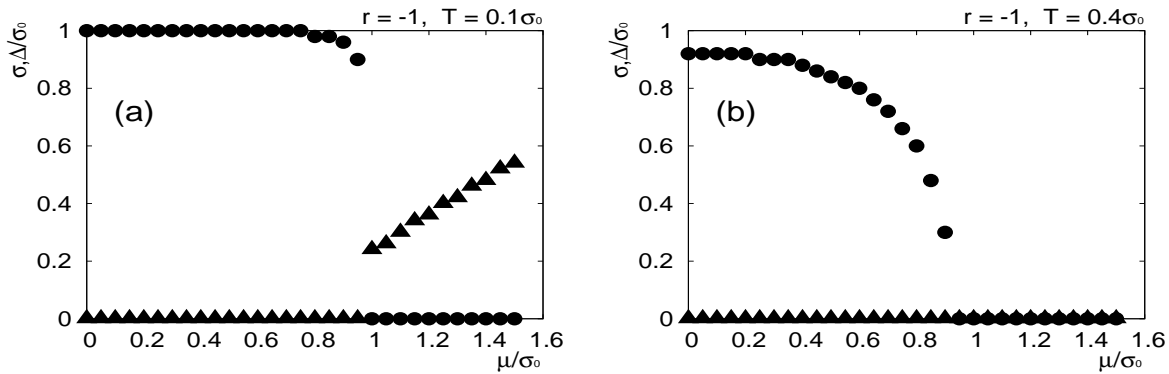}
\end{center}
\caption{\label{fig:Tcondensate}The two gaps $\sigma$ (circles) and $\Delta$ 
  (triangles) for ($r$, $T$) $=$ ($-1$, $0.1\sigma_0$) and ($-1$, $0.4\sigma_0$)}
\end{figure*}
Now we turn to the $T \neq 0$ case. We display 
the results for ($r$, $T$) $=$ ($-1$, $0.1\sigma_0$) and ($-1$, $0.4\sigma_0$)
in Fig.~\ref{fig:Tcondensate}. 
These panels show that the $\bar{q}q$ condensate for $\mu=0$
is $1.0\sigma_0$ at $T=0.1\sigma_0$ and $0.92\sigma_0$ at $T=0.4\sigma_0$.
The $qq$ condensate for $\mu = 1.0\sigma_0$ is 
$0.24\sigma_0$ at $T=0.1\sigma_0$ and $0$ at $T=0.4\sigma_0$.
Thus, as $T$ increases,
the $\bar{q}q$ and $qq$ condensates decrease.
Note that the $qq$ condensate appears at $T=0.1\sigma_0$ and does not appear at
$T=0.4\sigma_0$, which indicates that $\Delta$ disappears at
high temperature.
More detailed analysis shows that the $qq$ condensate at $\mu = 1.0\sigma_0$
disappears for $T > 0.15\sigma_0$ (see Sec.V).
It should be noted that, within our numerical accuracy,
there is no region where the $\bar{q}q$ and $qq$ condensates coexist, also
at finite temperature.
The results for other values of $r$ are qualitatively the same.
As the temperature increase, the condensates become smaller and 
completely disappear at the critical temperature.
This is the signal of the phase
transition from the condensate state to the normal state.

From Fig.~\ref{fig:Tcondensate}, we see that the phase transition for 
$T = 0.1\sigma_0$ is apparently of the 
first order and the case for $T = 0.4\sigma_0$, it is of the second order.
Investigating the thermodynamic potential as a function of the $\bar{q}q$
condensate as in \cite{Ulli}, 
we have confirmed this fact.
We will discuss the phase transition
and its order in the next section in more detail.

\section{The phase diagram}
\begin{figure*}
\begin{center}
\includegraphics[width=17cm,height=17cm]{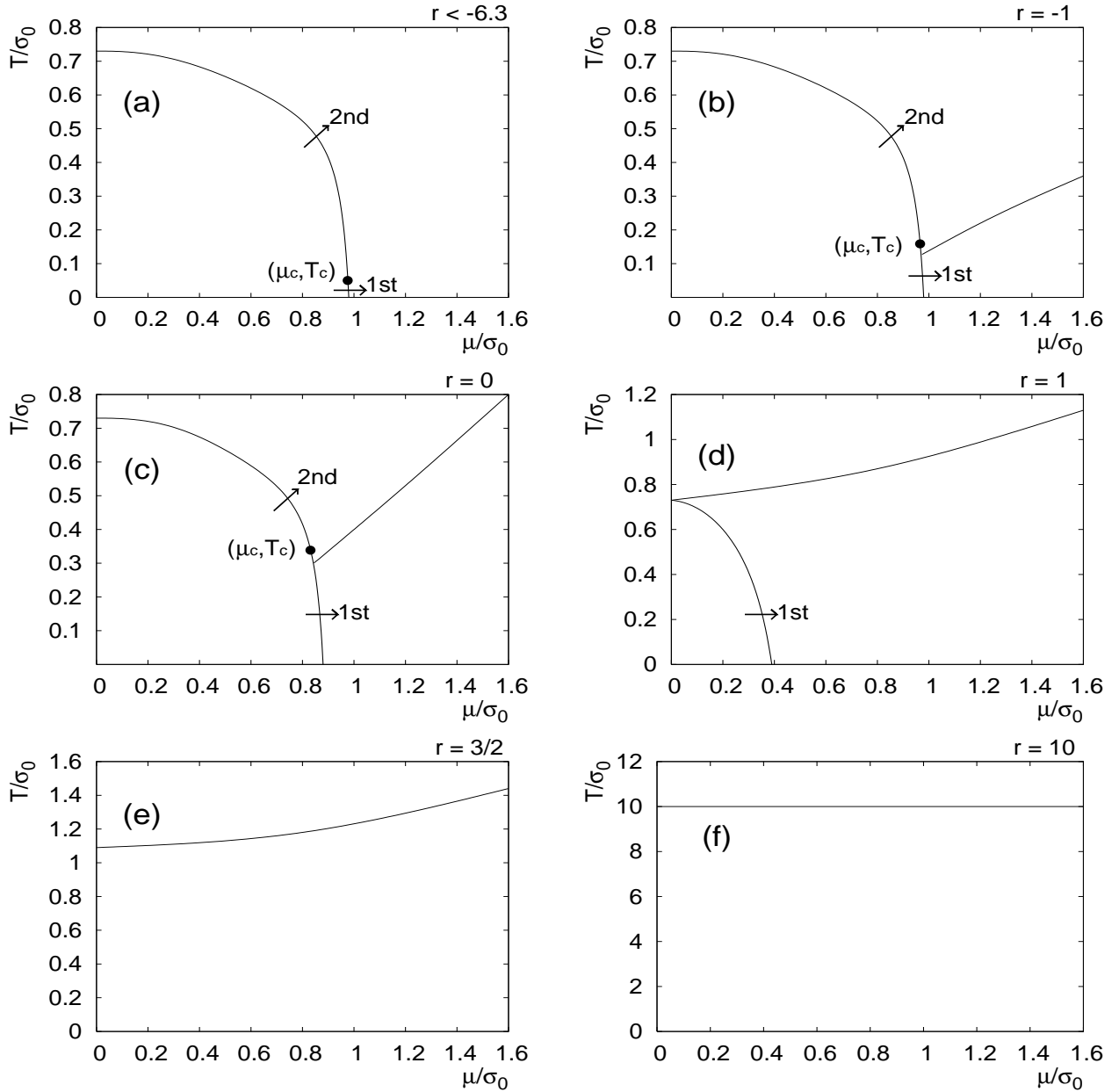}
\end{center}
\caption{\label{fig:phase}The phase diagram of the 3D GN model.}
\end{figure*}

Through minimizing the thermodynamic potential, one obtains the phase diagram.
In Fig.~\ref{fig:phase}, we display the phase diagrams for various values of $r$.
For $r < -6.3$, there appears the pure $\bar{q}q$ condensate phase at low temperature
and density and no $qq$ condensate phase appears.
In the cases
of $r= -1,\, 0$, the phase diagrams bear resemblance to that of QCD. 
For $r=-1$ with $T=0$, 
the phase transition from the $\bar{q}q$ condensate phase to the $qq$ condensate phase 
takes place at $\mu \simeq 0.98\sigma_0$. 
For $\mu = 0$, as $T$ increases, 
the transition from the $\bar{q}q$ condensate phase to the normal 
phase takes place at $T \simeq 0.73\sigma_0$, which applies also for 
$r= -10,\, 0, \,1$. On the other hand, in the cases of $r=3/2$ and $10$, the 
transition temperature for $\mu=0$ are $T \simeq 1.08\sigma_0$ and $10\sigma_0$,
respectively.

As $r$ increases, the region of 
$\bar{q}q$ condensate phase shrinks toward the $\mu$ axis and the region of the $qq$ 
condensate phase increases toward the $T$ axis. For $r = 3/2$ and $10$, the $\bar{q}q$ 
condensate does not exist only and the $qq$ condensate appears.
Through numerical analysis, we have found that 
the $\bar{q}q$ condensate phase disappears completely 
at $r \simeq 1.15$.

The points ($\mu_c,T_c$) shown in the panel (a), (b) and (c) in Fig.~\ref{fig:phase} 
represent the critical points from
the first order phase transition to the second order. 
The phase transition below the critical temperature $T_c$ is the first order and 
above $T_c$ is the second order. On the other hand in the panel (d), 
there is no critical point and 
the phase transition from $\bar{q}q$ condensate to the $qq$ condensate is always 
the first order.
The more detailed analysis tells us that the critical point disappears
for $r \simeq 1$, and it always appears between 
the $\bar{q}q$ phase and the normal phase. 
With respect to the $qq$ condensate, the phase transition from the $qq$ to the normal phase
is always of the second order.

\section{Summary and conclusions}
We have studied the $\bar{q}q$ and $qq$ condensates in the 3D GN model with 2d spinor
quarks,
and obtained the phase diagram  for various values of $r=\Delta_0/\sigma_0$.

We have found that the behaviors of the
$\bar{q}q$ and $qq$
condensates at $T=0$, in Fig.~\ref{fig:0condensate},
bear resemblance to that of the 4D NJL model \cite{Mei}:
With increasing $r$,
the $qq$ condensate becomes more dominant and the $\bar{q}q$ 
condensate disappears completely
for $r > 1.15$. 

Fig.~\ref{fig:0condensate} and Fig.~\ref{fig:Tcondensate}
show that, both for $T=0$ and $T \neq 0$,
there is no region where the $\bar{q}q$ and $qq$ condensates coexist.
This is a characteristic feature 
in the 3D GN model which does not happen in the 4D NJL model.

From Fig.~\ref{fig:phase} (b) and (c), we see that for $r=-1,\,0$,
there is a close resemblance between 
the phase diagrams for $r = -1,\,0$ and that of QCD.
The $\bar{q}q$ condensate phase here corresponds to 
hadronic phase in QCD, and the $qq$ condensate phase corresponds to color 
superconducting phase. In the case for $r<-6.3$, there does not appear
the $qq$ condensate phase, and the phase diagram shows close similarity
with the QCD phase diagram without color superconducting phase.
However the diagrams for $r=1 ,\, 3/2$ and $10$ are very different from the QCD case.
Especially when $r = 3/2$ and $10$, there is no $\bar{q}q$ condensate and only the 
$qq$ condensate exists.

The circles in Fig.~\ref{fig:phase} indicate the 
critical points with respect to the $\bar{q}q$ phase transition from first order
to the second order.
Note that the phase transition from the $\bar{q}q$ condensate to the $qq$ 
condensate is always of the first order. As seen from Fig.~\ref{fig:0condensate} and
\ref{fig:Tcondensate},
when $qq$ condensate
arises, the $\bar{q}q$ condensate disappears rapidly.

We have found that the phase structure drastically changes according
to the value $r=\Delta_0/\sigma_0$. With increasing $r$, the $qq$ condensate
becomes larger. This means that the $qq$ condensate becomes more dominant when
$\Delta_0$ increases. On the other hand, $\Delta_0$ is related to the $qq$ coupling 
constant $G_D$ through
Eq.(\ref{delta0}), and as $\Delta_0$ increases, $G_D$ increases.
In the same reason, when $G_S$ increases (decreases), $\sigma_0$
becomes large (small), which causes $r$ to decrease (increase).
With these observations in mind, we can conclude that when $G_D$ is small, the
$\bar{q}q$ condensate is dominant over the $qq$ condensate, and the $qq$ condensate
becomes larger when $G_D$ increases.
This is the same phenomenon as seen in the 4D NJL model.

Finally it should be mentioned again that we have found the absence of the coexisting phase
in the 3D GN model {\it with 2d spinor quarks}. As mentioned in Sec.I, the 3D GN model
with 4d spinor quarks bears resemblance to the 4D NJL model. Then it is worth studying
the phase diagram in the 3D GN model with 4d spinor representation to see
whether the coexisting phase appears as in the 4D NJL model or does not appear
as in the present case.

\begin{acknowledgments}
I would like to express my sincere gratitude to A. Niegawa
and M. Inui for useful discussions.
\end{acknowledgments}

\appendix

\section{The derivation of Eq.(\ref{det})}
The determinant of $G^{-1}$ can be rewritten as
\begin{eqnarray}
    \mbox{Det} G^{-1}  = &\mbox{Det}& \left(
    \begin{array}{cc}
      (\slash{p} - \sigma + \mu \gamma^0) {\bf 1}_f \tilde{{\bf 1}}_c
    & i \epsilon \tau_2 \Delta {\bf 1}_s \\
      i \epsilon \tau_2 \Delta^{*} {\bf 1}_s
    & (\slash{p} - \sigma - \mu \gamma^0) {\bf 1}_f \tilde{{\bf 1}}_c
    \end{array}\right) \nonumber \\
    &\times &\mbox{Det}
    \left(\begin{array}{cc}
    (\slash{p} - \sigma + \mu \gamma^0) {\bf 1}_f & 0   \\
    0 & (\slash{p} - \sigma - \mu \gamma^0) {\bf 1}_f
    \end{array}\right),
    \label{detsep}
\end{eqnarray}
where $\tilde{{\bf 1}}_c$ is the unit matrix and $\epsilon$ is the antisymmetric
matrix in color (red and green) space, $\epsilon_{rg}= -\epsilon_{gr} =1$.
For a $2 \times 2$ block matrix with matrices $A$, $B$, $C$ and $D$, we have the 
identity
\begin{align}
    & \mbox{Det} \left( 
    \begin{array}{cc} 
    A \; & \, B \\ 
    C \; & \, D 
    \end{array} 
    \right) = \mbox{Det} \left( - C B + C A C^{- 1} D \right) \, .
\end{align}
Replacing $A$, $B$, $C$ and $D$ with corresponding elements in
the first line of Eq.(\ref{detsep}), we get
\begin{eqnarray}
    && \mbox{Det} \left(
    \begin{array}{cc}
    (\slash{p} - \sigma + \mu \gamma^0) {\bf 1}_f \tilde{{\bf 1}}_c
    & i \epsilon \tau_2 \Delta {\bf 1}_s \\
    i \epsilon \tau_2 \Delta^{*} {\bf 1}_s 
    & (\slash{p} - \sigma - \mu \gamma^0) {\bf 1}_f \tilde{{\bf 1}}_c
    \end{array}\right) \nonumber \\
    &=&
    \mbox{Det} \left(
    -|\Delta|^2 + p_0^2 - \vec{p}^{\,\, 2} + \sigma^2 -\mu^2 - 2\sigma \slash{p} - \mu \slash{p} \gamma^0
      + \mu \gamma^0 \slash{p}
    \right)^4 \nonumber \\
    &=& 
        \Bigl(p_0^2 
        - E^2 - \mu^2 - |\Delta|^2 
        - 2 \sqrt{E^2 \mu^2 + \sigma^2 |\Delta|^2}\Bigr)^4 \nonumber \\
        &&\times \Bigl(p_0^2 
        - E^2 - \mu^2 - |\Delta|^2 
        + 2 \sqrt{E^2 \mu^2 + \sigma^2 |\Delta|^2}\Bigr)^4,
\end{eqnarray}
which leads to the first two factors in Eq.(\ref{det}).
After calculating the second line of Eq.(\ref{detsep}), we finally obtain
$\Omega(\sigma,|\Delta|)$ in Eq.(\ref{det}).

\section{Renormalization in vacuum}

As mentioned in Sec.I,
the standard $O(N)$ 3D GN model is renormalizable in the leading $1/N$ order. 
This is also the case for the thermodynamic
potential Eq.(\ref{Thermody}) in the present model with in the mean field approximation.

Following the procedure as in \cite{Klimenko} (see also \cite{GN}),
we carry out the renormalization and obtain the renormalized thermodynamic potential.

The divergent part of the potential is $\Omega_0$ in Eq.(\ref{thermo03d}) and the
divergent integral is written as
\begin{align}
     \Omega_{0\,{\rm div}} & (\sigma,|\Delta|) = 
     -\frac{3}{\pi^2} \sigma^2 \Lambda - \frac{2}{\pi^2} |\Delta|^2 \Lambda
\label{Ozerodiv}
\end{align}
It is to be noted that $\Omega_{0\,{\rm div}}$ is independent of $\mu$.

To eliminate this divergence, we introduce the counter Lagrangian $\mathcal{L}_C$,
Eq.({\ref{counterL}}). 
For determining $Z_S$ in $\mathcal{L}_C$, we 
consider the one-loop radiative correction to the
$\sigma$ propagator in the $2+1$ Minkowski momentum space as calculated in
\cite{Klimenko},
\begin{align}
D_{\sigma}(p^2) &= \frac{-i}{(1/2G_S)+i\Pi(p^2)}, 
  \label{sigmaprop} \\
\Pi(p^2) &= - N_f N_c \int \! \frac{d^3 k}{(2\pi)^3} 
  \frac{{\rm Tr} [\slash{k}(\slash{k}-\slash{p})]}{k^2(k-p)^2},  
\label{oneloop} 
\end{align}
where $N_f$ and $N_c$ are the numbers of flavors and colors.
$\Pi(p^2)$ shown above is ultraviolet divergent, and the counterterm $Z_S$
is introduced so as to eliminate its divergent contribution. 
Here we employ the renormalization condition
\begin{equation}
D_{\sigma}(p^2) = -2i \,G_S  \, , \quad {\rm at}  \quad p_0^2-\vec{p}^{\,2}=-\alpha^2.
\end{equation}
This means that the counterterm $Z_S$ should satisfy the following relation
\begin{equation}
 Z_S = - i \Pi(p^2)\bigr|_{p_0^2-\vec{p}^{\,2}=-\alpha^2}.
 \label{repoint}
\end{equation}
Going into the Euclidean space, and restrict the integration region with a sphere of 
radius $\Lambda$. Then after evaluating 
integral Eq.(\ref{oneloop}) and using the condition Eq.(\ref{repoint}),
we obtain
\begin{equation}
 Z_S =  \frac{6}{\pi^2}\Lambda - \frac{3}{4}\alpha.
\end{equation}
In the similar manner, one obtains the renormalization constant $Z_D$:
\begin{equation}
 Z_D =  \frac{2}{\pi^2}\Lambda - \frac{1}{4}\alpha.
\end{equation}
Introduction of the counter Lagrangian $\mathcal{L}_C$ eliminates $\Omega_{0\, {\rm div}}$,
and we obtain the renormalized $\Omega_{0r}$.




\end{document}